\documentstyle[aaspp4]{article}
 
\lefthead{Elmegreen}
\righthead{Bulge Formation as a Maximum Starburst}
\slugcomment{scheduled for ApJ, vol. 517, May 20, 1999}
\begin{document}
 
\title{Galactic bulge formation as a maximum intensity starburst}
 
\author{Bruce G.~Elmegreen\altaffilmark{1}}
\altaffiltext{1}{IBM Research Division, T.J. Watson Research Center,
P.O. Box 218, Yorktown Heights, NY 10598}
 
\begin{abstract}
Properties of normal galactic star formation, including the density
dependence, threshold density, turbulent scaling relations, and 
clustering properties, are applied to 
the formation of galactic bulges.  One important difference is
that the bulge potential well is too
deep to have allowed self-regulation or blow-out
by the pressures from young stars, unlike galactic disks or
dwarf galaxies.  As a result, bulge formation should have been at the
maximum rate, which is such that most of the gas would get converted
into stars in only a few dynamical time scales, or
$\sim10^8$ years.  The gas accretion phase can be longer than this, but
once the critical density is reached, which depends primarily on the
total virial density from dark matter, the formation of stars in the
bulge should have been extremely rapid.  Such three-dimensional star
formation should also have formed many clusters, like normal disk star 
formation today.  Some of these clusters may have survived as 
old globulars, but most got dispersed, although they might still be
observable as concentrated streams in phase space.  

\end{abstract}

Subject headings: stars: formation -- galaxies: formation -- 
galaxies: starburst

\section{Introduction}
 
On a large scale, star formation in galaxies appears to be regulated by
the self-gravity of the gas, producing new stars at a rate proportional
to the average gas density $\rho$ multiplied by the standard
gravitational rate, $\left(G\rho\right)^{1/2}$.  This leads to the
Schmidt law of star formation (Madore 1981; Larson 1988;
Elmegreen 1991; Kennicutt
1998) and the observation of a decreasing gas consumption time at
higher densities, which is essentially the 
starburst phenomenon (Larson 1987; Elmegreen 1994;
Kennicutt 1998).

Large-scale star formation also has a density threshold
(Madore, van den Bergh, \& Rogstad 1974; Gallagher \& Hunter 1984;
Guiderdoni 1987),
below which the rate per unit gas mass is much less than the inverse
gravity time. This threshold has been attributed to several 
types of 
limits: (1) a minimum column density for strong self-gravity of a gas
disk in the presence of galactic differential rotation,
i.e., $\sigma_{crit}\sim\kappa a/\left(\pi G\right)$ (Toomre 1964;
Goldreich \& Lynden-Bell 1965; Quirk 1972; Fall
\& Efstathiou 1980; Zasov \& Simakov 1988; Kennicutt 1989; Elmegreen 1991), 
and (2) a minimum column
density corresponding to the minimum pressure (which scales as $\sim
G\sigma^2$) that gives a cool thermal gas phase,
$\sigma_{crit,th}=7\left(\sigma_g/\left[\theta\sigma_T\right]
\right)^{0.5}\left(I/I_\odot\right)^{0.6}$ M$_\odot$
pc$^{-2}$ for
radiation field $I$ relative to the local value, $I_\odot$, 
total gas+stellar column density $\sigma_T$, and ratio of thermal
to total gas pressure, $\theta$
(Elmegreen
\& Parravano 1994). The outer edges of galaxies 
where star formation slows abruptly
(Ferguson et al. 1998)
are probably the result of
contributions from both factors, considering that gaseous self-gravity
becomes low (Kennicutt 1989) and the HI becomes nearly
pure warm-phase (Braun 1997). In addition, the outer parts of galaxies
are beyond the realm of strong stellar density waves, 
particularly outside the outer Lindblad resonance
for the dominant pattern speed of spiral wave modes in
the disk (Bertin et al. 1989). The absence of stellar
density waves removes one of the primary mechanisms
for organizing the gas into star-forming clouds.

On smaller scales, there is a correspondence between the duration of star
formation in a region and the turbulent crossing time 
(Efremov \& Elmegreen 1998), which suggests that
self-gravity and turbulence together partition the gas into clouds and
cloud clumps, and that the clouds are rather transient, turning into
stars quickly before a chaotic turbulent boundary destroys the
favorable environment (e.g., see turbulence simulations in
Ballesteros-Paredes, V\'azquez-Semadeni \& Scalo 1998).  The
observation of similar initial stellar mass functions in clusters
(Massey 1998) and in integrated galactic light 
(Bresolin \& Kennicutt 1997) is
consistent with this turbulent time scale, because it implies that intermediate
and high mass stars are not the primary agents of destruction
for molecular clouds. 
Such destruction would steepen the combined IMF from all clouds
above the Salpeter
power-law slope of $\sim-1.35$ to a slope of $\sim-4$ (Elmegreen 1998,
1999), which is observed only rarely in remote regions of the field
(Massey et al.  1995).

Also consistent with this short time scale is the common observation
that young or embedded star clusters still contain significant
subclustering, presumably from the pre-stellar cloud.   All of this 
stellar substructure would have been mixed together 
if the cluster age were much more than a crossing time. 
Examples of young clusters with small-scale
subclustering are W33 (Beck et al. 1998), G 35.20-1.74 (Persi et al.
1997), IC 348 (Lada \& Lada 1995), NGC 2264 (Piche 1993), and many
other clusters in the Large Magellanic Clouds (Elson 1991).

These observations are useful for understanding the connection between
large and small scale star formation. Stars form in 
self-gravitating clouds on a wide range of  scales, but the density of
these clouds increases, and the gravitational time scale decreases,
for smaller scales, probably
because of the way turbulence structures the
gas. This structure is also hierarchical, so many
small regions of star formation come and go
while the large region enclosing them continues
to form stars.  As a result, {\it the large scale
controls the overall star formation time, while the smaller scales
determine the net efficiency}, i.e., the
fraction of the gas that is turned into stars
in all of the transient small scale events. 

The surface density threshold for 
star formation has been observed for a wide variety of regions,
including normal spiral galaxies (Kennicutt 1989; Caldwell et al.
1992), 
low surface brightness galaxies (van der Hulst
et al. 1993), starbursts (Shlosman
\& Begelman 1989; Kennicutt 1998) and embedded gas disks
in elliptical galaxies (Vader \& Vigroux 1991). 
However, in all cases considered
so far, the geometry has been either disk-like, as for large parts
of galaxies, or ring-like, as in some
nuclear starbursts (Elmegreen 1994). A threshold for star
formation in spheroidal systems like dwarf galaxies and galactic bulges
has not been seriously considered.  Nevertheless, the basic
principles of star formation that have been derived for disk
galaxies should also have
applications in other regions, including the galactic bulge, as long as
the three-dimensional geometry for self-gravity and various turbulent and
magnetic processes are properly considered.

Here we discuss theoretical and observational evidence that the
criterion for star formation underlying most of the galactic observations
is one in which the {\it volume} density of gas is comparable to or
larger than the total virial density in that region, including 
stars and dark matter. This threshold has already been suggested for
dwarf and irregular galaxies (Elmegreen et al. 1996). It also has
a natural extension to three dimensions, 
and therefore should be appropriate to the formation of galactic bulges.
In addition, we suggest that the likely
formation time of the bulge was only a few dynamical 
times, as it is in local molecular clouds, and that this short time leads to
bulge formation luminosities
that are extremely large, 
consistent with observations of galaxies
at high $z$ (Giavalisco et al. 1995, 1996;
Steidel et al. 1996). 
Moreover, the pressures resulting from star formation give an
interesting condition for self-regulation and
blow-out that depends only on the depth of
the potential well.  This condition is such that normal galaxy disks and
dwarf galaxies can regulate their star formation to remain near
or below 
the threshold, but
galactic bulges and whole elliptical galaxies cannot.

Another implication of using normal star formation
processes for galactic bulges is the
prediction that stars should form in numerous dense
clusters. The remnants of these clusters might be visible today as stellar
streams or other concentrations in phase space, and as bulge (and halo)
globulars. 

\section{A three-dimensional threshold for star formation}

Spitzer (1942) noted the existence of a maximum stable gas mass for a
given temperature in a fixed background potential from stars. This
maximum mass corresponds to a threshold ratio of the gas density to the
virial density. Above this density, gas pressure cannot support a cloud
against the combined gravity from itself and the background stars.

Some applications of this three-dimensional threshold density have already
been made for dwarf and irregular
galaxies. The sizes of the largest regions of star formation
(``star complexes'') in dwarf, irregular, and spiral galaxies 
spanning a factor of $10^4$ in luminosity were found to always
equal half the Jeans' length for a fixed gaseous velocity dispersion of
$\sim10$ km s$^{-1}$, provided the gas density is comparable to the background
virial density that is given by the Tully-Fisher relations between
galaxy size, dynamical velocity, and absolute magnitude (Elmegreen et
al. 1996). This implies that when the gas density exceeds the
virial density, star formation begins on a scale equal to the Jeans length.
These results are interesting here 
because dwarf galaxies are not
thin disks and they have no spiral density waves. Thus the
star formation threshold for dwarfs might be similar to that for other
three-dimensional structures. 

Spitzer's derivation was based on the virial theorem.
It gave the result that an isothermal gas cloud embedded in a
stellar spheroid is unstable if its central density $\rho_c$ exceeds
$\rho_{vir}(1+\beta/2)^{-1}$ for gaseous gravitational potential $\beta
GM_{gas}/R$.

We recalculate this result here by direct integration of the equations of
hydrostatic equilibrium for a mixture of cool gas and a dissipationless
fluid that might be considered to be stars, dark matter, or
both. Here we just call
the second component dark matter. The equations of equilibrium are
$dP_{gas}/dr=-\rho_{gas} g$ and $dP_{DM}/dr=-\rho_{DM} g$, respectively.  
The gravitational
acceleration comes from both components according to Poisson's equation
in spherical coordinates: $\nabla\bullet {\bf g}=4\pi G
\left(\rho_{gas}+\rho_{DM}\right)$. The results are normalized to the
central dark matter density, $\rho_{DM,0}$, the 
dark matter velocity dispersion,
$a_{DM}=\left(P_{DM}/\rho_{DM}\right)^{1/2}$, and the gravitational
length and mass scales given by these quantities, namely,
$a_{DM}/\left(4\pi G\rho_{DM,0}\right)^{1/2}$ and
$a_{DM}^3/\left(3\left[4\pi\right]^{1/2}G^{3/2}\rho_{DM,0}^{1/2}\right)$,
respectively.

The top left diagram of figure 1 plots various density profiles for
dark matter (dashed lines) and gas (solid lines), considering
different central gas densities. The lower dashed line corresponds to
the upper solid line. The top right diagram shows the total gas mass as
a function of the central gas density, for various ratios of the gaseous
velocity dispersion to the dark matter dispersion. When the gaseous 
velocity dispersion is small, the total gas mass peaks
at intermediate values of the central gas density. This is because at lower gas
density the gas sphere has a low density throughout a large size, and at
higher gas density, the gas sphere has a high density throughout a small
size. As the gaseous dispersion approaches the stellar dispersion,
however, the total mass of the gas sphere approaches infinity, as does the
mass of the stellar sphere, because of the usual infinite extent of
isothermal sphere solutions. 

The lower left diagram in figure 1 shows the peak total gas mass versus the
velocity dispersion ratio as a solid line, and the ratio of the total
gas mass to the dark matter mass inside the gas sphere as a
dashed line. The gas mass ratio is about $0.03$ at low dispersion ratio.
The final result is the gas density ratio, plotted in the lower right.
When the central gas density exceeds about 12.6 times the
central dark mass density, or the average gas density inside the
half-mass gas radius exceeds about 6 times the central dark matter density, 
there are no stable solutions which permit the addition of more mass to a cool 
gas component of a gas+dark matter galaxy. 
Any additional gas mass leads to catastrophic collapse of the whole gas sphere. 
If write the total gas+dark 
matter density as the virial density, then this result gives a
critical average gas density
\begin{equation}\rho_{crit}\sim\rho_{vir}.\end{equation}

This is about the same value as the critical gas density in spiral
disks. The disk condition, from Kennicutt (1989), is usually written in
terms of the surface density, \begin{equation}\sigma_{crit}=0.7{{\kappa
a_{gas}}\over{3.36 G}}. \end{equation} For a flat rotation curve with
speed $V$ and radius $R$, the epicyclic frequency $\kappa=2^{1/2}V/R$.
To get the virial density, we take $V^2=GM/R=4\pi G\rho_{vir}R^2/3$. The
surface density gets converted into a volume density using the scale
height $H$, as $\rho=\sigma\left(2H\right)^{-1}$. The scale height for
gas is about $H=a_{gas}\left(2\pi G\rho_T\right)^{-1/2}$, where $\rho_T$
is the total midplane density, including stars. Thus $
\rho_{crit,disk}
\sim0.8\left(\rho_{vir}\rho_T\right)^{1/2}$. Locally,
$\rho_{vir} \sim2.5\times10^{-24}$ g cm$^{-3}$ for $V=220$ km s$^{-1}$
and $R=8.5$ kpc; this is about $1/3$ the total midplane density, so we
take $\rho_T\sim3\rho_{vir}$ as representative. Then
\begin{equation}
\rho_{crit,disk}\sim1.3\rho_{vir}.\end{equation}

The virial density is a reasonable threshold for major events of
galactic star formation because the main processes that resist gaseous
self-gravity in disks, namely, shear and rotation, have effective time
scales (Oort $A^{-1}$ and $\kappa^{-1}$, respectively) that equal
approximately $\left(G\rho_{vir}\right)^{-1/2}$. Gaseous self-gravity,
on the other hand, has an effective time scale of
$\left(G\rho_{gas}\right)^{-1/2}$. Thus self-gravity dominates when
$\rho_{gas}>\rho_{vir}$ over a length scale larger than
the Jeans length. The interstellar magnetic field will not change
this result much if it is in equipartition with the turbulent motions,
because the first step in cloud formation following a gravitational
instability consists of motions along the magnetic field, and the field
pressure does not resist this motion much.  Spiral waves in 
disks may trigger substantial amounts of star
formation, even in sub-critical disks, but since the wave
incidence rate is on the order of a rotation rate, the net time scale
for star formation, averaged over a rotation, is still about the orbit
time (Wyse \& Silk 1989). In addition, spiral waves seem to require dense
gas for strong amplification (Lubow, Cowie, \& Balbus 1986), so again we get
the result that when the gas density is near the threshold for strong
self-gravity, at $\rho_{gas}\sim\rho_{vir}$, star formation, even in
spiral density waves, becomes rapid.

\section{The threshold star formation rate}

Considering the above discussion, the threshold star formation rate per
unit volume in any region will be about equal to the threshold gas
density, which is essentially the local virial density, divided by the
local dynamical time, or \begin{equation}
SFR\approx{{\epsilon\rho_{vir}}\over{t_{dyn}}}\sim \epsilon
G^{1/2}\rho_{vir}^{3/2}\sim{{\epsilon }\over{G t_{dyn}^3}},
\label{eq:sfr}\end{equation}
where $\epsilon\sim0.1$ is the fraction of the gas that is turned into
stars in each major star formation event (given by the
combined efficiencies of star
formation on smaller scales).  The dynamical time is $t_{dyn}\sim R/V$. For
$t_{dyn}= 10^7t_{dyn,7}$ years, this rate becomes \begin{equation} SFR
\approx {{200\epsilon}\over{t_{dyn,7}^3}}\;\;{\rm M}_\odot \;{\rm
kpc}^{-3}\;\;{\rm yr}^{-1}.\label{eq:sfrt}\end{equation}

Meurer et al. (1997) found an upper limit to the areal star formation
rate in starburst nuclear disks that is equal to $\sim45$ M$_\odot$
kpc$^{-2}$ yr$^{-1}$. This comes from equation \ref{eq:sfrt} if we
multiply the volume star formation rate by a galactic thickness of
$\sim0.3$ kpc, use $\epsilon\sim0.1$, and take a dynamical time
appropriate for a nuclear disk, $t_{dyn}=R/V\sim5$ My. Equation
\ref{eq:sfrt} predicts that even higher star formation rates will be
found in smaller regions where $t_{dyn}$ is smaller, although the
fraction of galaxies undergoing such intense bursts will be small
because of rapid gas consumption during the active phase. 

Meurer et al. (1997) also discussed 
wind limitations to star formation rates. 
They estimated the pressure from a star-forming region to be 
\begin{eqnarray}
P_{SF}\sim7.2\times10^{-11}
{\rm dyne\;cm}^{-2} \times\left(
{{SFR_{2D}}\over{{\rm M}_\odot\;{\rm kpc}^{-2}{\rm yr}^{-1}}}\right)\\
=1.1\times10^{7}
{\rm dyne\;cm}^{-2} \times\left(
{{SFR_{2D}}\over{{\rm gm}\;{\rm cm}^{-2}{\rm s}^{-1}}}\right).
\end{eqnarray}
This pressure is proportional to the two-dimensional star
formation rate, $SFR_{2D}$, which is the galactic thickness
$2H$ times the 3-dimensional rate from equation \ref{eq:sfr}.
Thus, at threshold (in cgs units),
\begin{equation}
P_{SF}\sim1.1\times10^7 \epsilon\left(G\rho_{vir}\right)^{1/2}\rho_{vir}2H.
\end{equation}

This disruptive pressure should be compared with 
the self-gravitational pressure that holds the region together,
which is proportional to the square
of the mass column density.  For a disk, 
$ P_{grav,disk}\sim \left(\pi/2\right)G\sigma_{gas}\sigma_{T}$
for total column density in the gas layer, $\sigma_T$. 
This equation follows from the use of $H^2=
a_{gas}^2/\left(2\pi G\rho_T\right)$
for gas disk scale height $H$ and from the relation $\sigma=2H\rho$; then
$P_{grav}=\rho_{gas}a_{gas}^2$, as expected.  If we take
$\rho_{gas}\sim\rho_{vir}$ at threshold, 
and $\rho_T\sim3\rho_{vir}$ as above, then
$P_{grav,disk}\sim6\pi G\rho_{vir}^2H^2.$
For a sphere, $P_{grav,sph}\sim\left(\pi/2\right)G\rho_T\rho_{gas}4H^2$
again if we define $H=a_{gas}/\left(2\pi G\rho_T\right)^{1/2}$ in the same way.
But now $\rho_{T}=\rho_{vir}$ 
for the problem considered
in figure 1, and $\rho_{gas}\sim\rho_{vir}$
inside the half-mass radius. Thus
$ P_{grav,sph}\sim2\pi G\rho_{vir}^2H^2.$
To cover both cases, we write this pressure as
\begin{equation}
P_{grav}=\alpha\pi G\rho_{vir}^2H^2,\end{equation}
where $\alpha=6$ and $\rho_T=3\rho_{vir}$ for a disk, and 
$\alpha=2$ and $\rho_T=\rho_{vir}$ for a sphere.  

Now we see that the star formation pressure exceeds the self-gravitational
binding pressure when 
\begin{equation}
1.1\times10^7\epsilon>\left(\pi/2\right)
\alpha\left(G\rho_{vir}\right)^{1/2}H\sim
1.7a_{gas};
\end{equation}
the coefficient 1.7 is the average between $\left(1.5\pi\right)^{1/2}=2.2$ 
for a disk
and $\left(0.5\pi\right)^{1/2}=1.2$ for a sphere. 
The coefficient on the left is in cgs units, namely, cm s$^{-1}$. 
Thus star formation pressure can blow apart a region at the
threshold density if the gaseous 
velocity dispersion, in km s$^{-1}$, is less than
a certain limit, 
\begin{equation}
a_{gas}<65\epsilon \;\;{\rm km}^{-1}\;{\rm s}^{-1}.
\end{equation}

This limit is surprisingly simple: for typical $\epsilon\sim0.1$, it
says that star formation can regulate or stop itself 
in a disk (e.g., Franco \& Shore 1984)
or dwarf galaxy (e.g., Dekel \& Silk 1986) where the gaseous
velocity dispersion is low, $a_{gas}\sim7$ km s$^{-1}$, 
but it cannot regulate or stop 
itself in a young galactic bulge or elliptical galaxy, where
the velocity dispersion from the potential well 
is large, $a_{gas}\ge100$ km s$^{-1}$.  Star
formation might also fail to regulate itself in nuclear disks or
interacting galaxies, which have large gaseous velocity dispersions
too (Elmegreen et al. 1993; 
Irwin 1994; Elmegreen et al. 1995; Brinks et al. 1997; Kaufman et al. 1997). 
Such failure to self-regulate should lead to 
rapid and intense star formation on a time scale of 
only a few dynamical times.

Rapid star formation in bulges and elliptical systems also leads to large
luminosities.  We convert the star formation rate from equation 
\ref{eq:sfrt} to a volume emissivity
$j_{SF}$, following Meurer et al. (1997), and get
\begin{equation}
j_{SF}\sim{{9\times10^{11}\epsilon}\over{t_{dyn,7}^3}}
\;{\rm L}_\odot\;{\rm kpc}^{-3}.
\end{equation}
This expression may be used to 
get the total luminosity by multiplying it by the volume, $4\pi R^3/3$
for radius $R$, and then the
result can be written as a function of the dynamical speed, $V$,
using $t_{dyn}=R/V$.
The resultant luminosity is
\begin{equation}
L\sim4\times10^{12}\epsilon\left({{V}\over{100\;{\rm km}\;{\rm s}^{-1}}}
\right)^3\;{\rm L}_\odot.\end{equation}

Evidently, the threshold starburst luminosity also 
depends only on the depth of the potential well of the 
galaxy. 
This result should apply
to many types of systems, since we have assumed very little about the
nature of the star formation or how the gas got into the potential well. 
For merging galaxies, the dynamical speed can be $V\sim200$ km s$^{-1}$
or more, leading to star-formation luminosities
of $L\sim10^{12}$ L$_\odot$ for times of $t_{dyn}\sim10^8$ years
($t_{dyn}/\epsilon\sim10^9$ years if the residual gas from
inefficient star formation continues to make stars). 
For bulges in which $V\sim100$ km s$^{-1}$, 
the luminosity during bulge formation can be 
$L\sim10^{11}$ L$_\odot$  
for a total formation time
of $t_{dyn}/\epsilon\sim 10R/V\sim10^8$ years. 
These values are 
consistent with observations of the Lyman $\alpha$ emitting galaxy
at $z=3.4$ found by Giavalisco et al. (1995), who suggest they are
witnessing the formation of a galactic bulge or elliptical galaxy
(see also Lilly 1999).
According to this theory, the formation
of a typical galactic bulge could be all over within
$\sim100$ million years once it starts, i.e., within
$\sim t_{dyn}/\epsilon$ for a complete conversion
of gas into stars. This result is consistent with recent observations
of rapid bulge formation in the Milky Way (Renzini 1999). 

The presence of a critical density for star formation 
in three-dimensional systems implies that there should have been a delay
in the onset of bulge formation during part of the accretion and cooling
phase of the galactic gas near the center. 
Star formation had to wait until the accretion built up a gas
density in the center comparable to
$\sim10\times$ the density of dark matter there. 
After this critical density was reached, star formation should have
been extremely rapid, producing $10^{11}$ L$_\odot$ for $10^8$
years and then dying off (depending on the rate
of continued gas accretion). 
The bulge should not have been able to regulate
this rapid star formation by supernovae, 
winds or other young stellar pressures, because the total binding pressure
exceeded the star formation pressure. Any winds that were generated had
to move between the dense gas clumps and find their way out of the
bulge in streams 
without completely expelling these clumps from the system. 

This starburst model of bulge formation differs from 
the model by Wada, Habe \& Sofue (1995), who suggested that
a nuclear starburst ejects a giant gas bubble, which forms stars and
relaxes into a bulge. It also differs from other bulge formation
models in which the bulge stars are ejected individually from the
disk as a result of bar-driven or other resonances (Combes et al. 
1990; Pfenniger \& Friedli 1991;
Friedli \& Benz 1995), or ejected in bulk when a bar dissolves
(Hasan, Pfenniger, \& Norman 1993;
Friedli \& Benz 1993).
In the present model, the bulge formed in a three-dimensional
distribution (Eggen, Lynden-Bell, and Sandage 1962)
as it does in the clusters of local molecular clouds,
except on a much larger scale.  A prediction of this
model can then be made by analogy to all other known star formation: 
{\it the bulge should have been born with an extremely clumpy distribution of
young stars}, i.e., composed almost entirely of clusters with a hierarchical
distribution and a 
mass function something like
$M^{-2}dM$ (Fleck 1996; Elmegreen \& Falgarone 1996). 
Presumably some of these clusters remain today as globular clusters (Surdin 1979;
Okazaki \& Tosa 1995; Elmegreen \& Efremov 1997), but most should have been
dissolved by interactions and variable tidal forces 
(Gnedin \& Ostriker 1997). In that case, they may still be observable
as stellar streams, which are concentrations in phase space. 
This was apparently the fate of local disk star clusters that were
dispersed with age (Eggen 1989; Dehnen 1998).

\newpage
\begin{figure}
\vspace{6.3in}
\includegraphics{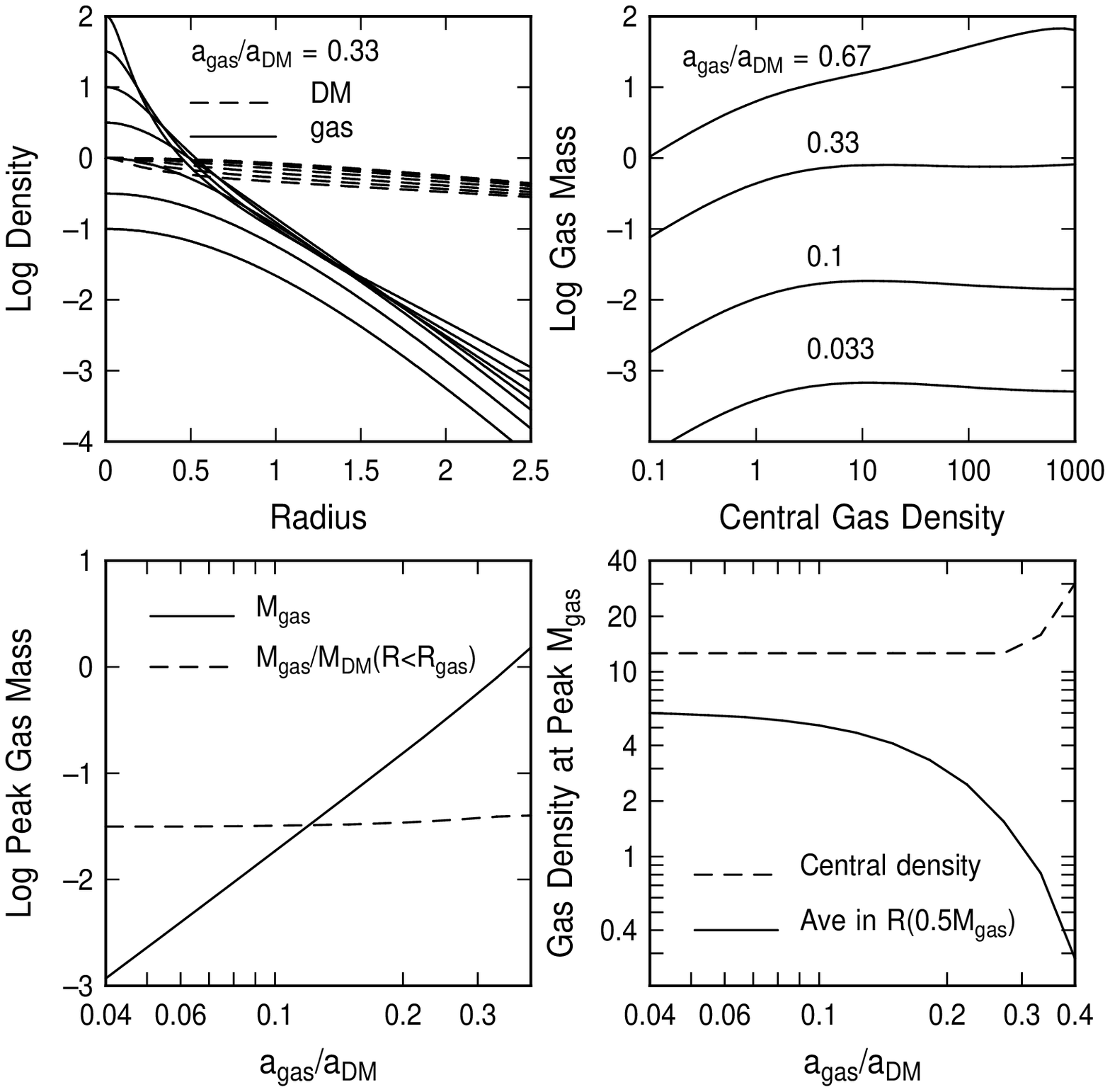}
\caption{(top left) Density profiles for gas and dark matter
in a spherical potential, considering different central gas
densities. (top right) Gas mass versus central density for
four values of the ratio of the gas to the dark matter
velocity dispersion. (lower left) Maximum gas mass, 
and ratio of the maximum gas mass to the dark matter mass in the 
gas sphere, versus velocity dispersion ratio. (lower right)
Central gas density at the maximum stable gas mass (solid line)
and average gas density inside the radius containing half the
total gas mass (dashed line), versus velocity dispersion ratio. }
\label{fig:scalo}
\end{figure}

\end{document}